\documentclass[cits]{PoS}

\title{The impact of localized overlap eigenmodes on RMT measurements and topology}

\ShortTitle{The impact of localized overlap eigenmodes on RMT measurements and topology}

\author{\speaker{Anna Hasenfratz}\\
        Department of Physics, University of Colorado, Boulder, CO-80309-390\\
        E-mail: \email{anna@eotvos.colorado.edu}}

\author{Roland Hoffmann\\
        Department of Physics, University of Colorado, Boulder, CO-80309-390\\
        E-mail: \email{hoffmann@pizero.colorado.edu}}

\author{Stefan Schaefer\\
        NIC, DESY, Platanenallee 6, D-15738 Zeuthen, Germany\\
        E-mail: \email{stefan.schaefer@desy.de}}

\abstract{The low energy eigenmodes of the continuum QCD Dirac operator 
           are extended, but on the lattice, due to discretization effects, the Dirac operator can have localized
eigenmodes. These non-physical modes can introduce strong lattice artifacts for observables that are sensitive to chiral symmetry, especially in mixed action simulations. We study how these lattice artifacts depend on the parameters of the overlap operator and their affect on the distribution on the Dirac eigenmodes and the topological susceptibility.
          }

\FullConference{The XXV International Symposium on Lattice Field Theory\\
		 July 30-4 August 2007\\
		 Regensburg, Germany}

\begin{document}

\section{Introduction}

Mixed action simulations combine the advantages of chiral operators
in the measurement with relatively fast configuration generation,
but their success to a large extent depends on how close the valence
and sea quark actions are. At the very least one requires that
the valence action does
not introduce any new (large) lattice artifacts. Our goal is to find
the {}``best'' overlap action, in the sense of smallest lattice artifacts within a group of simple
actions, to use with our ongoing Wilson action
dynamical simulation\cite{Hasenfratz:2007rf,Hasenfratz:2007dc,Roland-proc}.
In a mixed
action situation the valence operator interacts with the sea quarks
through the vacuum, therefore it is best to match the valence fermions
to the dynamical ones via physical quantities that are sensitive
to the vacuum but independent of the valence quark mass. We have considered
two such observables, the topological susceptibility and the
distribution of the  infrared eigenmodes of the valence
Dirac operator\cite{Hasenfratz:2007iv}. 

In the phase where chiral symmetry is spontaneously broken the low energy eigenmodes
of the Dirac operator are expected to be extended, delocalized. The
lattice Dirac operator can have many localized eigenmodes, but as
long as these modes remain separate form the low energy infrared modes,
they do not affect physical predictions. We have investigated the
localization properties of the eigenmodes of the Wilson Dirac operator
and several different overlap operators. We found that while both
the Wilson and overlap operators have localized eigenmodes, these
modes do not mix with the infrared modes of the Wilson Dirac operator,
but can become part of the low energy spectrum of the overlap operator.
The density of these non-physical modes depend on the parameters of
the overlap construction, on the gauge configurations and on the lattice
spacing. We can relate these modes to the localized modes of the kernel
operator and argue that they are due to dislocations of the gauge
configurations. 
These non-perturbative lattice artifacts can strongly affect chiral observables in mixed action simulations.

Our observation suggests that in order to minimize scaling violations
in overlap simulations it is not sufficient to rely on perturbative
$O(a)$ improvement but that non-perturbative lattice artifacts due
to dislocations also have to be considered.

\section{Notations and parameters}

While this work is motivated by dynamical simulations with nHYP smeared
improved Wilson fermions\cite{Hasenfratz:2007rf,Hasenfratz:2007dc,Roland-proc},
the lattice artifacts of the overlap valence operators can be equally studied
on quenched configurations. We used about 1000 $12^{4}$ configurations
generated with Wilson plaquette action at $\beta=5.8458$ ($a\approx0.12$fm). 

Our definition of the massless overlap operator is \begin{equation}
D_{{\rm ov}}=R_{0}\left(1+d(d^{\dagger}d)^{-1/2}\right)\;,\quad d=D_{K}-R_{0}\;,\label{eq:Overlap_def}\end{equation}
 where $D_{K}$ is the kernel operator and $R_{0}$ denotes the center
of the overlap projection. We choose $D_{K}$ to be the Wilson operator
with nHYP smeared gauge connections \cite{Hasenfratz:2001hp,Hasenfratz:2007rf},
both unimproved and with tree level ($c_{\,{\rm SW}}=1$) clover improvement. 

The choice of the parameter $R_{0}$ in the overlap construction is
rather arbitrary, as long as it is larger than the eigenvalues of
the physical, infrared modes of the kernel operator but smaller than
the doubler modes, and the resulting overlap operator is local. Since
the infrared edge of the spectrum, $\lambda_{{\rm crit}}$, varies
with the kernel operator, the quantity $\Delta R_{0}=R_{0}-\lambda_{{\rm {\rm crit}}}$
characterizes the overlap operator better than $R_{0}$ itself. We
have chosen two different $\Delta R_{0}$ values with both of our
action, $R_{0}=1.0$ ($\Delta R_{0}\approx0.7$) and $R_{0}=0.7$
($\Delta R_{0}\approx0.4$) with the unimproved $c_{{\rm SW}}=0$
action and $R_{0}=1.0$ ($\Delta R_{0}\approx0.92$) and $R_{0}=0.3$
($\Delta R_{0}\approx0.22$) with the improved $c_{{\rm SW}}=1.0$
action. All four actions lead to a local overlap operator.

\section{The eigenvalue spectrum of the kernel and overlap Dirac operators}

\begin{figure}
\centering\includegraphics[height=7cm]{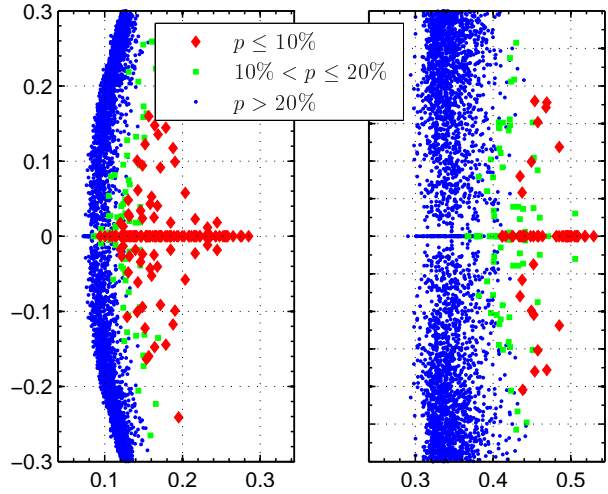}
                                                                                 
\caption{The spectrum of the two kernel operators used in this study. Both
are nHYP smeared Wilson operators, one with tree level $c_{\,{\rm SW}}=1$
clover coefficient (left panel), the other with $c_{\,{\rm SW}}=0$
(right panel). The different plot symbols correspond to different
localization levels of the corresponding eigenvectors . \label{fig:Dirac-spectrum}}
\end{figure}

First we consider the eigenvalue spectrum of the kernel operators.
Figure \ref{fig:Dirac-spectrum} shows the 40 lowest magnitude eigenvalues
on 100 configurations with both the $c_{\,{\rm SW}}=1$ and $c_{\,{\rm SW}}=0$
kernel actions. The spectrum of the nHYP smeared $c_{\,{\rm SW}}=1$
operator appears much more chiral than the unimproved one, its eigenvalues
are concentrated around a unit circle. This is what makes this action
appealing in dynamical simulations. 

A simple and very intuitive measure of the localization of the eigenmodes
is the participation number or inverse of the inverse participation
ratio IPR \cite{Gattringer:2001mn} \begin{eqnarray}
p & = & IPR^{-1}\nonumber \\
IPR & = & V\sum_{x}|\psi(x)|^{4}\,,\label{eq:IPR}\end{eqnarray}
 where $\psi(x)$ is the normalized eigenvector of the Dirac operator.
In Figure \ref{fig:Dirac-spectrum} the different plotting symbols
correspond to different participation numbers of the eigenmodes, and
one observes a strong correlation between $p$ and $\Delta R$, the distance from
the outer edge of the circle. Toward the center of the eigenvalue
circle all modes appear to be localized with small $p$ for both actions.
However the spectrum of the clover improved action has many more localized
modes in the vicinity of the physical, IR range. 

The participation number is only a qualitative measure: while a very
small $p$ certainly implies a localized mode, a large value does
not necessarily mean a coherent extended one. Finite volume analysis
can distinguish the localized and extended modes. If the typical eigenmodes
in a given region of the eigenvalue circle 
correspond to extended modes, their average participation
$\bar{p}$ should be volume independent, while in the region where
most of the eigenmodes are localized $\bar{p}$ will decrease with
the inverse of the volume. Comparing data on $12^{4}$ and $16^{4}$
lattices we found constant $\bar{p}$ values for $\Delta R<0.03$ and
$1/V$ dependence for $\Delta R\ge0.05$ for the $c_{\, SW}=1$ spectrum.
This finite volume analysis suggests that on the $12^{4}$ lattices
eigenmodes with participation number $p<0.40$ at $\Delta R\approx0.05$
are already localized. 

The overlap construction {}``projects'' all the modes to the Ginsparg-Wilson
circle. %
\begin{figure}
\centering\includegraphics[height=5.5cm]{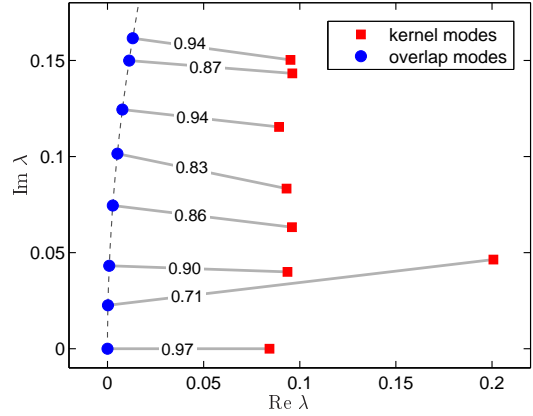}
                                                                                 
\caption{Low energy eigenvalues of the kernel and overlap operator on a configuration
where the overlap operator has a localized IR mode. The lines connect
the overlap and kernel modes with the highest overlap, and the magnitude
of the inner product of the modes is also shown. \label{fig:Comp} }
\end{figure}
A typical example of how the overlap construction transforms the kernel
modes is shown in Figure \ref{fig:Comp}, where we compare the eigenvalues
of the $c_{\, SW}=1$ kernel operator and the corresponding overlap
operator.
All but one of the kernel eigenmodes shown are
extended with large participation numbers, the only exception is the
mode in the inner part of the circle that has $p\approx0.04$. The
eigenmodes of the overlap operator are also extended with one exception,
the mode with the lowest imaginary value has $p\approx0.08$. The
extended overlap eigenmodes  connect strongly to a kernel mode,
with overlap between the wave functions of 80\% or larger as indicated
in the figure. The extended, near infra-red eigenmodes change little
under the overlap projection, their eigenvalues basically move 
straight out to the Ginsparg-Wilson circle. The localized mode, on
the other hand, behaves differently. 
Both the overlap and kernel wave functions are sharply concentrated, they couple mainly
to a small instanton or dislocation. The overlap of the wave functions
is sizable, \textasciitilde{}70\%, but the eigenvalues are quite different.
The overlap eigenvalue is small, the most infrared among the eigenmodes.
In general localized modes tend to stay localized under the overlap
projection, their overlap eigenvalue is frequently small, without
modifying the eigenvalues of the extended modes. 

\begin{figure}
\centering\includegraphics[height=5.5cm]{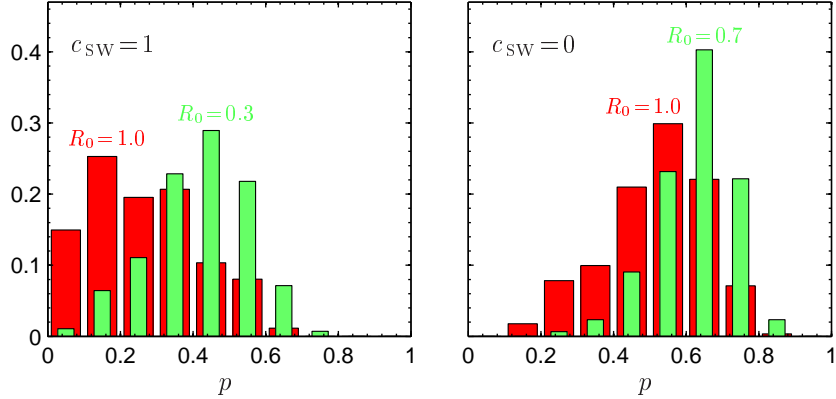}

\caption{The distribution of the participation number of the first non-zero
overlap eigenmodes in the $\nu=1$ sector, normalized by the number
of configurations. \label{fig:IPR}}
\end{figure}

To quantify the observations from Figures \ref{fig:Dirac-spectrum}
and \ref{fig:Comp} we have measured the participation number of the
low eigenmodes of our four overlap operators. Figure \ref{fig:IPR}
shows the distribution for the first \textit{non-zero} modes in the
$\nu=1$ topological sector. The result supports what we have expected
based on the eigenmodes of the kernel operator. The 
$c_{\,{\rm SW}}=1$ improved action kernel operator with $R_{0}=1.0$
has a lot of localized modes - possibly more than 50\% of the first
eigenmodes are localized. The other actions are considerably better.
When $R_{0}=0.3$ is used even with the clover improved action, many
of the localized modes are already to the right of the overlap center
and projected to the ultra-violet. Setting the clover coefficient
$c_{\,{\rm SW}}=0$ has a similar effect. Even with $R_{0}=1.0$,
corresponding to $\Delta R_{0}=0.70$, there are only a few localized
modes, and their number drops even further when $R_{0}=0.7$ ($\Delta R_{0}=0.40$)
is chosen.

\section{Consequences of localized overlap eigenmodes}

The distribution of the low lying Dirac eigenmodes should follow the
universal predictions of Random Matrix Theory for the extended eigenmodes
of the quenched systems if the volume is large enough, but localized
modes embedded in the IR can spoil the agreement.
According to RMT the probability distribution
of a given  eigenvalue of the Dirac operator in a fixed topological
sector  is a universal function and for quenched systems depends
only on one free parameter, $\Sigma V/a$, where $\Sigma$ is the
chiral condensate \cite{Damgaard:2000ah}. 

\begin{figure}
\centering\includegraphics[height=5.5cm]{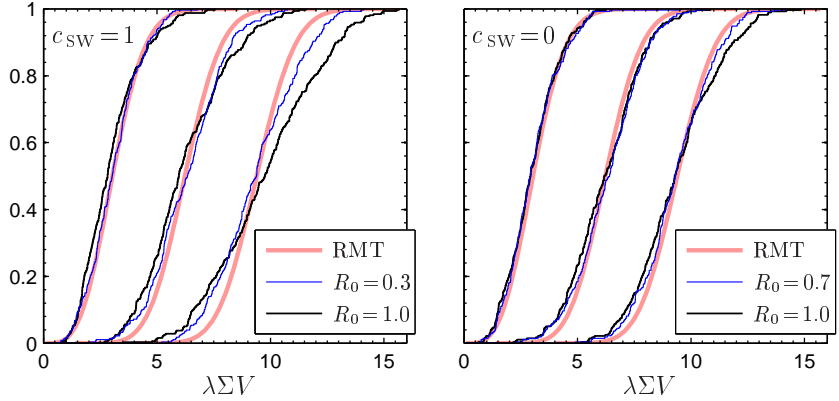}

\caption{The cumulative distribution of the first three eigenmodes in the
$\nu=1$ topological sectors. Left panel: $c_{\,{\rm SW}}=1$; right
panel: $c_{\,{\rm SW}}=0$. The smooth thick lines are the RMT predictions.
\label{fig:cumulative-distribution}}
\end{figure}

In Figure \ref{fig:cumulative-distribution} we present our results
for the cumulative (integrated) distribution using the four different Dirac operators
and compare them to the RMT predictions. 
As is evident from Figure \ref{fig:cumulative-distribution}, the first
eigenmodes are well described by RMT, but the agreement
gets progressively worse for the higher modes. In general the $c_{\,{\rm SW}}=1.0$
operators are worse than the unimproved ones. While   
the  $c_{\,{\rm SW}}=1.0$ smeared kernel action has much better chiral properties and the corresponding
overlap operator is also more local than with the unimproved kernel,  
it also has many more localized eigenmodes in the IR overlap spectrum.
These eigenmodes influence the distribution of the Dirac eigenmodes and ruin
the agreement with the analytical predictions.
\begin{figure}
\centering\includegraphics[height=5.5cm]{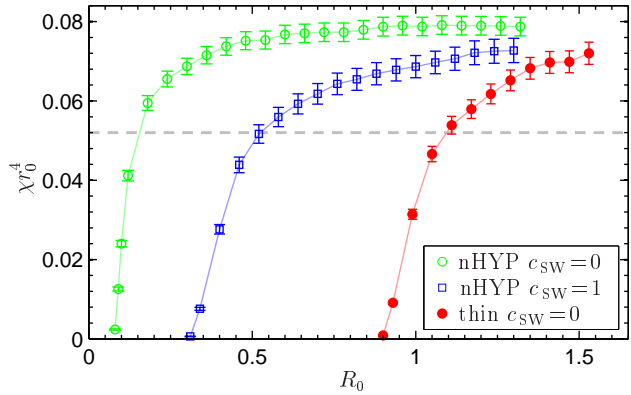}

\caption{The topological susceptibility as a function of the overlap parameter
$R_{0}$ with thin link and nHYP smeared overlap. The dashed horizontal
line is the continuum prediction from Ref. \cite{Durr:2006ky}. \label{fig:Topo_susc}}
\end{figure}

The topological susceptibility $\chi=\langle \nu^{2}\rangle/V$ is
defined via the index of the overlap operator. 
It is identical to the sum of the chirality
($\pm1$) of the real modes of the kernel operator up to $\lambda<R_{0}$
\cite{Niedermayer:1998bi}. The real modes of the kernel operator
are easiest to identify by measuring the eigenvalues of the Hermitian
operator $\gamma_{5}D_{K}$  and identifying when an eigenmodes crosses zero
\cite{Edwards:1998wx,DelDebbio:2003rn}.
In Figure \ref{fig:Topo_susc} we show the dimensionless quantity
$\chi r_{0}^{4}$ as a function of $R_{0}$ for the nHYP smeared
$c_{\,{\rm SW}}=0$ and 1 kernel overlap actions, and also for the
unimproved thin link kernel overlap action. To set the scale we use
$r_{0}/a=4.032$ from Ref. \cite{Necco:2001xg}. 
Recent studies of the topological
susceptibility using a pure gauge $F\tilde{F}$ topological charge
operator predict $\chi r_{0}^{4}=0.0524(13)$ \cite{Durr:2006ky},
while calculations with a thin link overlap operator give $\chi r_{0}^{4}=0.059(3)$
\cite{DelDebbio:2004ns} in the continuum limit. In Figure \ref{fig:Topo_susc}
we observe not only large cut-off effects, but strong dependence on
the $R_{0}$ parameter, especially for the $c_{\,{\rm SW}}=0$ actions.
This is the consequence of the large number of real
eigenmodes toward the center of the eigenvalue circle seen in Figure
\ref{fig:Dirac-spectrum}. Most of these modes are lattice artifacts,
dislocations. Overlap operators with smaller $R_{0}$ values are less
sensitive to these inner modes and therefore show smaller lattice
artifacts.

Comparing results we obtained for the topological susceptibility 
and for the eigenvalue distributions we observe
that lattice artifacts, or deviation form the continuum, correlate
closely for both observables with the density of localized overlap eigenmodes (Fig. \ref{fig:IPR}).

\section{Conclusion and Discussion}

We have investigated the lattice artifacts of different overlap operators
in quenched systems as reflected by the topological susceptibility
and the distribution of the low energy eigenmodes. We related the
observed cut-off effects to the existence of localized low energy
eigenmodes in the overlap spectra. 

While these localized modes, due to lattice dislocations, are lattice
artifacts and will go away in the continuum limit, their presence
can cause significant scaling violations at finite lattice spacing.
One can minimize these  by choosing a better
kernel operator, like the nHYP smeared operator we considered here,
and by tuning the $R_{0}$ parameter of the overlap construction as
small as the locality of the overlap operator would allow. 

In this paper we considered only quenched systems, but mixed action
simulations suffer from the same problem. 
Fully dynamical overlap simulations should fare better as there the
localized eigenmodes are suppressed just like any other small eigenvalue
mode, so while they are present, their number is at least not inflated.
Nevertheless an overlap operator that has small lattice artifacts
in quenched should also have smaller lattice artifacts in dynamical
simulations.

\section{Acknowledgment}
This work was partially supported by the US Department of Energy.


{\renewcommand{\baselinestretch}{0.86}
\bibliography{lattice}
\bibliographystyle{JHEP-2}}
 
\end{document}